# Detonation along laser generated micropinch for fast ignition

F. Winterberg

University of Nevada, Reno

**Abstract**

The proposed fast ignition of highly compressed deuterium-tritium (DT) targets by petawatt lasers requires energy of about 100kJ. To lower the power of the laser, it is proposed to accomplish fast ignition with two lasers, one with lower power in the infrared, and a second one with high power in the visible to ultraviolet region. The infrared laser of lower power shall by its radiation pressure drive a large current in a less than solid density plasma placed inside a capillary, while the second high power-shorter wave length-laser shall ignite at one end of the capillary a magnetic field supported thermonuclear detonation wave in a blanket made from solid DT along the outer surface of the capillary. The other end of the capillary, together with its DT blanket, is stuck in the DT target, where following the compression of the target the detonation wave ignites the target.

1. **<u>Introduction</u>**

The idea of fast ignition, invented by E. Teller for the ignition of large thermonuclear explosive devices was decades later proposed to be used for the ignition by petawatt lasers, of small highly compressed thermonuclear assemblies, with the patawatt laser, replacing the fission trigger in Teller's original concept [1, 2]. Before that, only the volume ignition of highly compressed targets was considered. The main advantage of the fast ignition concept is that by separating the compression from ignition much larger thermonuclear gains can be expected. And it also has the potential of pure DD (deuterium) burn, igniting a small amount of DT from where a deflagration into D can proceed. The main obstacle working against this concept is that the energy the petawatt laser must deliver is about 100 kJ.

One idea to overcome this problem called impact ignition, proposed by Murakami and Nagatomo [3], is to ablatively accelerate a small speck of dense matter (macron) to a high velocity igniting the target upon impact. To reach upon impact the ignition temperature of the DT reaction the macron must be accelerated to a velocity of the 1000 km/s [4]. The advantage of this concept is that it permits a comparatively slow accumulation of the kinetic macron energy over the distance it is accelerated. If the acceleration is done ablatively, the same laser used for compressing the target can simultaneously be used for the macron acceleration. In the scheme proposed by Murakami and Nagatomo, the macron is a small flyer plate placed inside a cone, with the cone stuck in the DT target. Experimental test of this idea [5], have achieved a flyer plate velocity of ~ 600 km/s, still short of the required 1000 km/s.

To reach even higher velocities it was proposed to drive the macron with the force of a burning magnetized DT plasma [6]. Going beyond this idea it is here proposed to ignite the compressed DT target by a small, magnetic field supported, thermonuclear detonation wave outside a micropinch inside a capillary, with one end of the capillary stuck into the target.

**2. Description of the micropinch detonation fast ignition concept**

The idea is described in Fig. 1:

1. A capillary C filled with a less than solid state density plasma ($n < n_0$) is surrounded by a solid state ($n = n_0 = 5 \times 10^{22}$ cm$^{-3}$) DT blanket.
2. An infrared laser pulse is injected from the left into the left side end of the capillary C. The radiation pressure of the infrared laser pulse accelerates the electrons inside the capillary to the right, generating a large electron current in that same direction, with the return current flowing outside the DT containing blanket.

3. A powerful shorter wave length laser pulse heats up a section of the blanket at its one end to the ignition temperature of the DT reaction, with the section larger than the Larmor radius of the charged DT fusion products (α-particles).

4. Provided the magnetic field set up by the electron current inside the capillary is large enough (of the order $10^7$ A), the DT fusion reaction α-particles are entrapped inside the DT blanket, and for that reason can launch a thermonuclear detonation wave D propagating outside the surface of the capillary to the right. If the other end of the capillary, together with its DT blanket, is stuck into the target, it can trigger in the target a thermonuclear deflagration.

## 3. The electric current set up in the capillary

This problem has been treated by the author in the context to compress by magnetic pressure fissile material to high densities to reduce the critical mass down to a small amount [7].

A coherent light beam with an average electric field $(\overline{E}^2)^{1/2}$ produces an electron drift motion of the plasma electrons with the drift velocity given by [8, 9]

$$v_d = e^2 \overline{E}^2 / m^2 c \omega^2 \qquad (1)$$

where ω is the circular frequency of the laser light, e and m are the charge and mass of an electron, c the velocity of light. For (1) to be valid $v_d \ll c$. For the laser radiation to be able to drive the electrons, the laser light frequency must be larger than the plasma frequency, i.e. $\omega \gg \omega_p$, where $\omega_p = (4\pi n e^2/m)^{1/2}$ with n the electron number density. If, for example, $n = n_0 = 5 \times 10^{22}$ cm$^{-3}$, valid for solid hydrogen, $\nu_p = \omega_p/2\pi = 2\times 10^{15} \, s^{-1}$, which corresponds to a wave length in the far ultraviolet.

Introducing the Poynting vector

$$S = (c/4\pi)\bar{E}^2 = P/A \qquad (2)$$

where P is the power of the laser beam in erg/s, and A, in $cm^2$, the area on which the laser beam is projected. With (2) one can write for (1)

$$v_d = (e/mc\omega)^2 4\pi P/A \qquad (3)$$

If the plasma in the capillary is hydrogen plasma or singly ionized plasma, the total current induced by the laser beam is

$$I = nev_d A \qquad (4a)$$

$$= c\left(\frac{\omega_p}{\omega}\right)^2 \frac{P}{I_A} \qquad (4b)$$

where $I_A = mc^3/e$ is the Alfvén current. In practical units it is equal to 17000 A. assuming that $\omega \approx \omega_p$, one can write for (4b)

$$P = I(I_A/c) \qquad (5)$$

For a current equal to $I = 10^7$ A, where the magnetic field is large enough to entrap the DT fusion reaction α-particles, one finds that $P \approx 10^{12}$ watt.

**4. Launching a detonation wave along the capillary**

For a current of $I = 10^7$ A, the Larmor radius $r_L$ of the charged DT fusion reaction products (α-particles) is about 1/10 the radius of the capillary, sufficiently short to entrap the α-particles in the DT reaction burning zone, which is the condition for detonation along the capillary in the DT containing blanket. For the ignition of this magnetic supported detonation, a volume of the order $A\sqrt{A} \approx 10^{-3}$ $cm^3$ has to be heated up to a temperature of $10^8$ $^0K$, the DT ignition temperature, in less than the Lawson time $\tau_L = 10^{14}/n_0 = 2\times10^{-9}$ sec. To heat a volume $10^{-3}$ cm$^3$ of solid DT to $10^8$ $^0K$ requires about $10^2$kJ of laser energy, at a power of $5\times10^3$ Watt, well within the reach of existing laser technology.

## 5. Incorporation of the proposed concept into the Kodama-Murakami-Nagatomo configuration

The evolution of the novel fast ignition concept, beginning with the Kodama-Murakami-Nagatomo configuration, is shown in Fig. 2.

The initial breakthrough of the petawatt laser fast ignition concept made by Kodama et al., is illustrated in Fig. 2a. Inserting a metallic cone into the DT target, aiding the petawatt laser pulse to reach the center of the compressed DT target, greatly increased the neutron yield. Because a much larger yield with this configuration seems possible only with a much larger petawatt laser, it was proposed by Murakami and Nagatomo to reach the same with about 10 times less laser power, utilizing the kinetic energy accumulation of a projectile ablatively accelerated to high velocities by the same laser.

Still better, seems to be an approach where energy is fed into the "ignitor" by thermonuclear reactions in a magnetized plasma at lower densities, ultimately this would have to be done by a small magnetic field supported thermonuclear detonation, as described in this configuration.

## Acknowledgement

This work has been supported in part by the U. S. Department of Energy under Grant No. DE-FG02-06ER54900.

**References**

[1]  N. G. Basov, et al. J. Sov. Laser Res. **13**, 396 (1992).

[2]  M. Tabak et al, Phys. Plasmas **1**, 1626 (1994).

[3]  M. Murakami and H. Nagatomo, Nucl. Instr. And Methods in Physics Research A 544, 67 (2005).

[4]  F. Winterberg, Z. Naturforsch. **19a**, 231 (1964).

[5]  M. Murakami et al., Plasma Phys. and Control. Fusion **47**, B815 (2005).

[6]  F. Winterberg, Phys. Plasmas **13**, 112702 (2006).

[7]  F. Winterberg, Z. Naturforsch. **28a**, 900 (1973).

[8]  J. H. Eberle and A. Sleeper, Phys. Rev. **176**, 1570 (1968).

[9]  O. S. Lieu and E. H. Shin, Appl. Phys. Lett. **20**, 511 (1972).

**Figure captions**

Figure 1: Detonation along capillary for fast ignition: $L_0$ longer –, $L_1$ shorter-wave length laser pulse; $n = n_0 = 5 \times 10^{22}$ cm$^{-3}$, solid DT; C capillary, I current inside capillary, B magnetic field outside capillary; DW detonation wave front, V fast ignition volume for target.

Figure 2: Evolution of the Kodama et al. (K), Murakami et al. (M) and "Capillary" (C) fast ignition configuration.

**Figures**

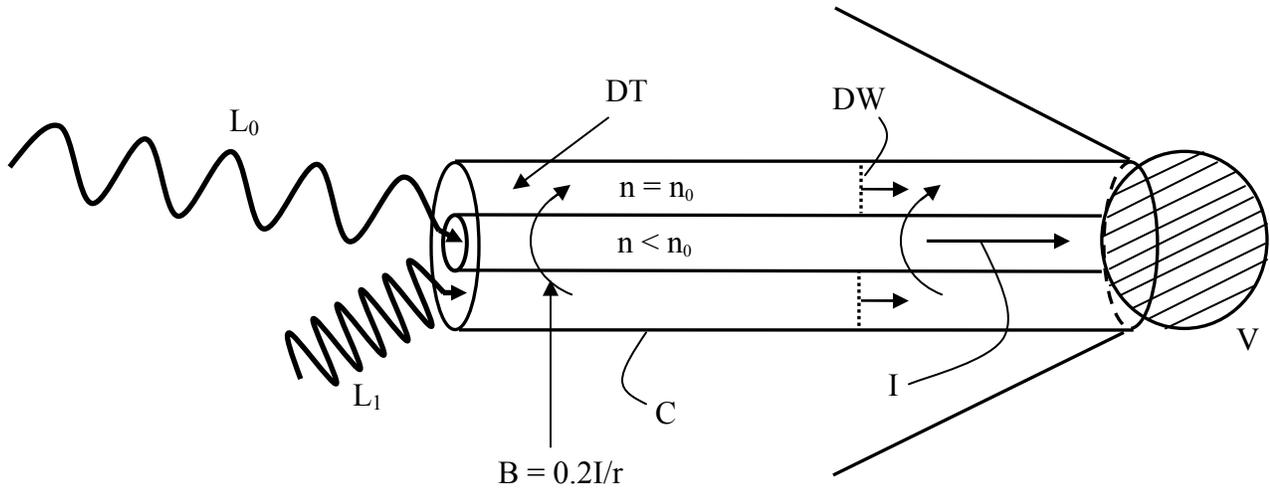

Figure 1

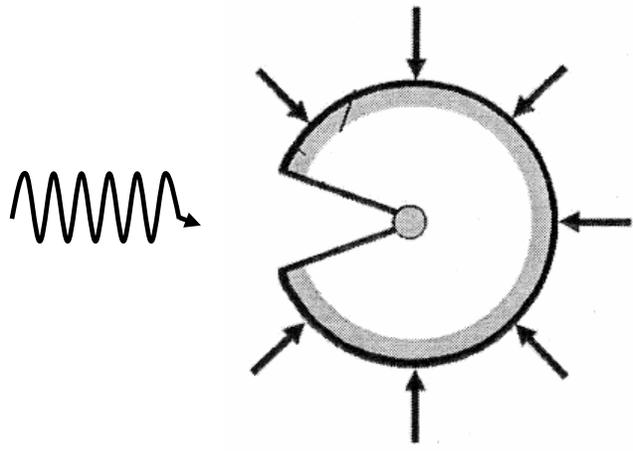

(2. a)

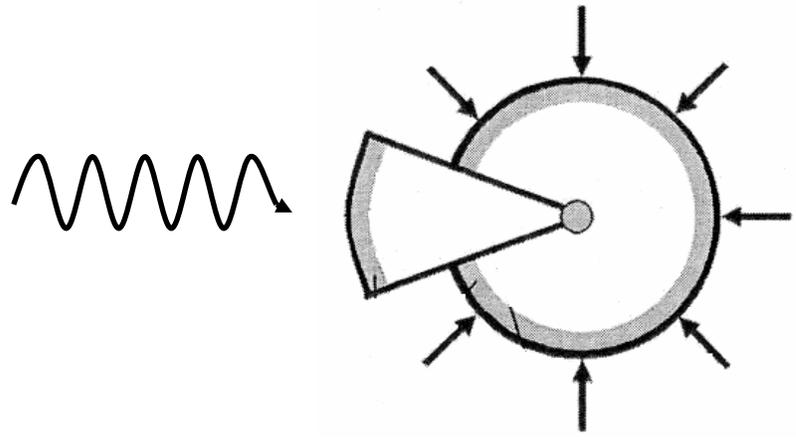

(2. b)

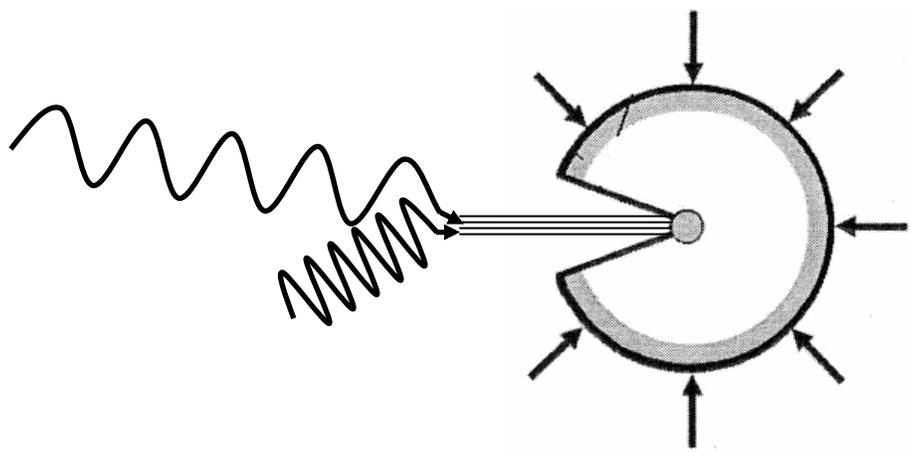

(2. c)

Figure 2

(K)

(M)

(C)